\documentclass{aastex}
\begin{document}

\title{In-orbit performance of the space telescope  NINA \\ and GCR flux measurements}

\author{V. Bidoli, A. Canestro, M. Casolino, M. P. De
Pascale,  G. Furano, A. Iannucci, A. Morselli,  P. Picozza, E. Reali, R. Sparvoli\altaffilmark{1
}}
\affil{Univ. of Rome ``Tor Vergata'' and INFN section of
Roma2, Via della Ricerca Scientifica 1, I-00133 Rome, Italy}
\author{A.  Bakaldin, A. Galper, S. Koldashov, M. Korotkov, A. Leonov, V. Mikhailov,
A. Murashov, S. Voronov}
\affil{Moscow Engineering Physics Institute, Kashirskoe Shosse 31, 115409 Moscow, Russia}
\author{M. Boezio, V. Bonvicini, R. Cirami, A. Vacchi, N. Zampa}
\affil{Univ. of Trieste and INFN
section of Trieste, Via A. Valerio 2, I-34147 Trieste, Italy}
\author{M. Ambriola, R.
Bellotti, F. Cafagna, F. Ciacio, M. Circella, C. De Marzo}
\affil{Univ. of Bari and INFN section of Bari, Via Amendola 173, I-70126 Bari, Italy}
\author{O.  Adriani, P.
Papini, S. Piccardi, P. Spillantini, S. Straulino}
\affil{Univ. of Florence and INFN section of Florence, Largo Enrico Fermi 2, I-50125 Florence, Italy}
\author{S. Bartalucci, G. Mazzenga, M. Ricci}
\affil{INFN Laboratori Nazionali di Frascati,  Via Enrico Fermi 40,  I-00044 Frascati, Italy}
\and
\author{G. Castellini}
\affil{Istituto di Ricerca Onde Elettromagnetiche CNR, Via Panciatichi 64, I-50127 Florence, Italy }

\altaffiltext{1}{Dept. of Physics, Univ. of Rome ``Tor Vergata'' and
INFN - Via della Ricerca
Scientifica 1,  00133 Rome, Italy. Email: Sparvoli@roma2.infn.it. Tel: +39-06-72594575 - Fax: +39-06-72594647}

\begin{abstract}
The NINA apparatus, on board the Russian satellite Resurs-01 n.4, has been
 in polar orbit since 1998 July 10,  at an altitude of 840 km.
Its  main scientific task  is to study   the  galactic,  solar
and anomalous
components of  cosmic rays in the energy interval 10--200 MeV n$^{-1}$. \\

In this paper we present a description of the instrument and its
basic operating modes. Measurements of Galactic Cosmic Ray spectra
will also be shown.
\end{abstract}

\keywords{Cosmic rays, Isotope Composition, Energy Spectrum,
Satellite, Silicon Detector}

\normalsize

\newpage

\section{Introduction}

With the launch of the telescope NINA on 1998 July 10 a wide
program of satellite cosmic ray observations began. The program
aims to study the cosmic ray radiation in a broad energy spectrum
(from 10 to 10$^5$ MeV n$^{-1}$) using a serious of dedicated
satellite missions.

NINA (a New Instrument
for
Nuclear Analysis) has been developed by a joint program of the Italian National
 Institute of Nuclear Physics (INFN) and the Moscow State Engineering and
Physics Institute (MEPhI).  INFN consists of several Italian
Institutes and Universities (WiZard group) who have carried out,
together with European and American partners, balloon-borne
experiments for the detection of cosmic antiparticles
 since 1989  \cite{pal8,pal7,pal6,pal5,pal4,pal3,pal2,pal1}.

The link with the Russian counterpart was established  in 1994, when the two sides
  started a collaboration and
conceived the Russian-Italian Missions (RIM), of which NINA  is the first step.

NINA's goal is to detect cosmic ray nuclei of galactic, solar and
anomalous origin, at 1 AU, from hydrogen to iron, between 10 and
200 MeV n$^{-1}$. The experiment is carried out on board  the
satellite Resurs-01 n.4, developed by the Russian space company
VNIIEM. The spacecraft was  launched into a  polar
sun-synchronous orbit of altitude 840 km  \cite{slc1,smi}.

NINA has been joined in space by a twin detector (NINA-2), placed
again in a  polar orbit but at a lower altitude (450 km). NINA-2 is housed  on board the Italian satellite MITA
\cite{marcoicrc}, launched on 2000 July  15 from the Plesetsk  launch facility in Russia by means of a Cosmos launcher. This second  mission
is intended to last for three years.

 The RIM missions will then continue
with the deployment of the PAMELA magnet spectrometer, which will
be installed onboard the satellite Resurs 01 n.5 and put in
orbit  at the beginning of the year 2003. The main objective of
PAMELA is to perform high precision measurements of antiparticle
spectra (positrons and antiprotons) in the energy range from 0.1
GeV up to  200 GeV. In addition, it will measure electrons,
protons and the nuclear components of cosmic rays, and  will
search for cosmic antinuclei \cite{pamela1,pamela}.

This article  reports on the NINA mission, its scientific tasks,
 the organization of the detector with its ancillary
instruments,  the interface with the satellite,  the launch
phase, and finally   the performance of the telescope in flight.
In addition,  it presents the reconstruction of the energy
spectrum of the galactic component of  $^4$He, $^{12}$C, and
$^{16}$O in solar quiet conditions.

\section{Scientific overview}

NINA has been built in order to investigate  the nuclear and
isotopic composition of low energy cosmic particles. Its
low-altitude polar
  orbit (about 1.1 Earth-radii) is particularly suitable for
performing   observations of particles of different  origin while traversing regions of  different
 geomagnetic cut-off.
The Earth's magnetic field is utilized
 as a spectrometer.  According to the coordinates  along the orbit where the particles
are detected, it is possible to make inferences about  their origin
which can be  galactic, solar, or anomalous.

\[Galactic \,\,Cosmic \, \,Rays\]

Galactic Cosmic Rays (GCR)  are a directly accessible sample of matter coming from
outside the Solar System.
The GCR energy spectrum  can be well represented by a power-law
energy distribution for energies above 1 GeV n$^{-1}$, but at
 lower energy
shows a strong attenuation due to the interaction between the Solar Wind and the cosmic particles
\cite{gal1}.  This is one of the reasons why GCR investigations below 200 MeV n$^{-1}$
have been relatively scarce in the past.

NINA started its mission in a period of medium solar activity; the next solar maximum
 is foreseen for the year 2000. Due to its technical characteristics and  its good energy, mass and angular resolution, the telescope is particularly suited
for exploring the low energy component of the cosmic radiation.
 The detector can record GCRs of very low energy (from 10 up to 200 MeV n$^{-1}$) in the
polar sectors of the orbit, where  geomagnetic effects are
virtually negligible.

\[Solar\,\, Energetic\,\, Particles\]

The most complete measurements of elemental abundances in the solar corona come from
measurements of high energy particles accelerated in the large Solar Energetic Particle (SEP)
events.

Initially it was thought that the energetic particles in  large
SEP events were accelerated in solar flares. In recent years,
however, it has become clear that in the large {\it gradual}
events particles are accelerated at shock waves that are driven
out from the Sun by Coronal Mass Ejections (CMEs)
\cite{ream1,ream2,ream3,cliver,ream4}. These shocks accelerate
the ions of the chemical elements in a fairly equivalent manner.
In contrast,  particles  accelerated in {\it impulsive} solar
flares show specific elemental enhancements produced by resonant
wave-particle interactions during stochastic acceleration of the
ions from the flare plasma.

New spacecraft observations  can extend Solar Energetic Particle
measurements to heavier elements, to rarer elements and to
isotopes. NINA can perform SEP observations in the polar sectors
of the orbit. Its good mass discrimination can help in the
determination of their composition and therefore in the
comprehension of the sources and acceleration mechanisms involved.

\[Anomalous\, \,Cosmic\, \,Rays\]

Anomalous Cosmic Rays (ACRs) are a low-energy component of
interplanetary particles that include the elements H, He, C, N,
O, and Ar \cite{klecker,simp0}. They are now known to originate
from interstellar neutral particles that have been swept into the
heliosphere, ionized by solar UV or charge exchange with the
solar wind, convected into the outer heliosphere, and then
accelerated to energies of $\sim$ 10 MeV n$^{-1}$ or more
\cite{fisk}.  It is commonly assumed that the bulk of ACR
acceleration takes place at the solar wind termination shock
\cite{pesses}.

The observation of the anomalous N, O, and Ne ionic charge composition with the SAMPEX satellite \cite{sampex}
confirmed the theoretical predictions that ACRs are only partially charged; more precisely, singly charged ions dominate only at energies below $\sim$ 20 MeV n$^{-1}$ \cite{klecker2}, while at higher energies multiply charged
ions become more abundant.

Being only partially ionized, ACRs have a much greater magnetic rigidity (at a given energy per nucleon) than either GCRs, which are essentially fully stripped, or SEPs, which have charge states characteristic of coronal temperatures. As a result, ACRs ca
n be observed to much lower invariant latitude with a polar orbiting spacecraft like NINA.

\section{The NINA instrument }

NINA consists of the following
 4 subsystems: a)
the {\it detector} (box $D1$), composed of 32 silicon layers
 and the
electronics for  signal processing, b) the {\it on-board computer}
(box $D2$), a dual microprocessor dedicated to data processing
and to the selection of the trigger and the acquisition mode
configuration, c) the {\it interface computer} (box $E$), which
rearranges the data coming
 from  box $D2$  and delivers them to the satellite telemetry system, and d) the
{\it power supply} (box $P$), which distributes the power supply
  to the different subsystems.

The weight and   electric power of the complete telescope   are
respectively  40  kg and 40 W, in accordance with the constraints
imposed by the satellite. To safeguard from possible malfunctions
and breaks, all electronic systems are  global  redundant.

\vspace{1cm}

 The  {\it detector} of NINA (box $D1$), manufactured by the Italian company Laben, is composed by 16  silicon planes.
Every plane consists of a pair of n-type silicon detectors, 60$\times$60 mm$^2$,
each read out by 16 strips mounted back to back with orthogonal orientation,
 in order to measure the X and Y coordinates of the  particle.
The strip pitch is 3.6 mm.

The
 thickness of the first pair of detectors, composing the first plane of NINA, is  (150 $\pm$ 15) $\mu$m; all
 the others, instead, are (380 $\pm$ 15) $\mu$m thick, for a total thickness of
 11.7 mm of silicon in the whole detector. The indetermination in the total silicon
thickness comes from the process of manufacturing, and is greater
(in percentage) for the first two thinner layers.
In order to reduce to the minimum the thickness of dead area interposed
between the silicon detectors, a special ceramic (Al$_2$O$_3$)
frame,
  passing only under the lateral  strips 1 and 16 and 625 $\mu$m thick,  has been utilized.
 The role of the ceramic is
to sustain the single silicon  structures and connect them
mechanically and electronically, by means of 64 pins, to the
corresponding mother-boards. A photo of the box $D1$ is shown in
Figure
 1.

Each plane of the detector with its electronics is mounted on an
aluminum mother-board. The 16 planes are vertically stacked  (a
sketch of the box $D1$ is visible in Figure
 2). The
interplanar distance is  1.4 cm except for
 the first two planes which are
separated by 8.5 cm in order to improve the incident particles
trajectory determination. They define  the   angular aperture
 of the telescope, which is about 32 degrees.
The 16 planes are modular, so that mechanically and electronically
they are interchangeable. Below the 16 silicon planes other 4
modules, dedicated to the trigger electronics, silicon power
supply, analog-digital conversion, and FIFO, are placed.

The 20 plane structure is housed in a cylindrical aluminum vessel of
284 mm diameter and 480 mm height, filled up with nitrogen at 1.2
atm. The vessel  is  2 mm thick, except for a  window above the
first silicon plane  where it  is
 reduced to 300  $\mu$m (Figure 2). The top
part of the vessel is rounded, while the bottom part houses the connectors for the  interfaces with the other parts of the detector.

The lateral strips (n.1 and n.16) of every silicon layer are used
for the Anticoincidence System (AC);  they are read together  by
the same electronic channel, except for  those of plane 1 where
they are physically disconnected. A total number of 448 electronic
channels, out of the 512 available in the box $D1$, are used for
the particle track and energy information; the Anticoincidence
System data occupy an additional 30 channels, while the remaining
34 are used for housekeeping data (16 plane currents, 2
temperatures, 4 voltages, 1 threshold level, 11 ratemeters at
different depths of the telescope), which monitor the status of
the whole instrument.

 The signals produced by  the incoming particles in the silicon strips  are
first amplified and  shaped. Every plane of the telescope has two 16 channels
preamplifiers.
Data are then converted to the digital format by means of a 12
bit ADC, with a full scale of
 2800 mip  (1 mip being equivalent
to 30400 electrons or about 105 keV of released energy). The
resolution per channel is thus about   0.68 mip ch$^{-1}$,
equivalent to  0.07 MeV ch$^{-1}$.
There are two
independent lines going to two different ADC's, for redundancy reasons. Only
one is operating at a given moment.  Before the ADC there are  two gain
amplifiers that can be selected depending on the acquisition mode: the first provides
an amplification of a factor 32 (used only for noise tests) and the second of
a factor 1 (active for normal acquisitions).

After conversion by ADC, data (1024 bytes/event) are  sent through a FIFO
 to an 8
channel bus interface with the {\it on-board computer } (box
$D2$), built again by the company Laben. Here all tasks of event
processing are performed, before sending the data to  the {\it
interface computer} (box $E$) for mass memory storage.

 The core of the  box
$D2$  are two 8086
 microprocessors working
with a clock speed of  4 MHz. In normal conditions both of them
are operating in Master-Slave mode: the Master  microprocessor
receives the  event  from  box $D1$ and performs pedestal
suppression and data reduction tasks, while the Slave is used to
format the data, according to the
 acquisition mode, and send them to the subsystem box
$E$. It also selects  the trigger logic, implements the Second
Level trigger and interfaces most of the telecommands with the
silicon detector.

The {\it interface computer } (box $E$) represents the last step
of the NINA data processing before the records  are sent to the
satellite for transmission, via telemetry, to ground. Two
exemplars of box $E$, for redundancy, have been built, both
realized by the Russian company VNIIEM. Finally the {\it power
supply} subsystem (box $P$), made also by VNIIEM,
 has the function of electrically connecting the satellite
with its various subsystems. The primary tension comes from the
solar panels, and it is nominally 27 V (between 24 and 34 V).
Starting from this, the box $P$ provides three different tensions,
two for the analog part (+6V, -6V) and one for the digital (+5V),
totally independent.

Further details about the instrument and its performance during a
test-beam session can be found in  \cite{nina1,nina2}.

\subsection{Operating modes}

\noindent NINA can work in different operating conditions,
switched automatically or via telecommand, which
affect the
trigger system. In particular:

\begin{enumerate}
\item  Two thresholds for the energy deposited
in the single silicon layers have been  implemented: a {\it Low
Threshold} (L.T.), corresponding to 2.5 mip, and a  {\it High
Threshold} (H.T.), corresponding to 25 mip.

The level of the threshold is fixed by  telecommand. As an
alternative, a system of automatic switching of the threshold,
activated by telecommand,  switches automatically from Low to
High Threshold whenever the external rate raises above 10 Hz, to
prevent the memory being saturated.

\item
The  strips 1 and 16 of every silicon layer, except the plane
first, are used in the  Lateral Anticoincidence System.

The hardware Lateral Anticoincidence can be turned off by
telecommand,  for instance in case of a  malfunction of one of the
lateral strips. In this condition, a software veto system (part of
the on-line Second Level Trigger) selects only tracks not hitting
the lateral strips. This procedure ensures that the Lateral
Anticoincidence rejection is always effective.

\item  The planes 15 and 16 can be used as Bottom Anticoincidence. The
default  operating mode adopts  plane 16 but, in case of need,
plane 15 can be
 selected
by telecommand.

 The Bottom Anticoincidence can be totally removed by telecommand,
allowing the detection of particles crossing the whole apparatus.

\end{enumerate}

\noindent The main trigger of the acquisition system is the following:
\begin{eqnarray*}
TRG\, M1 = D_{1x}\times D_{1y} \times
((D_{2x} +  D_{2y})  +
(D_{3x} +  D_{3y})) \,\,,
\end{eqnarray*}
where $D_{ij}$ denotes a signal above-threshold  coming from  plane
$i$, along view
$j$ ($j$=x,y). The logic OR of  planes 2 and 3
provides redundancy in case of a   failure of plane 2.

In the default operating mode, this trigger is used together with
the Lateral and Bottom Anticoincidence ON, in order to ensure the
complete containment of the particle inside the detector. This is
the condition which allows the best energy and nuclear
discrimination to be obtained by  NINA. Moreover, TRG M1  can be
used  with Low or High Threshold, defining two different
intervals of  nuclei which can be detected. In particular, TRG M1
in  High  Threshold mode removes  most of the protons from the
trigger. This is the most frequent configuration adopted in orbit.

\vspace{0.5cm}
It is possible to switch, via telecommand, to a
second trigger:
 \begin{eqnarray*}
TRG \, M2= (D_{2x} +  D_{2y})  \times
(D_{3x}  +
 D_{3y})  \times
(D_{4x}  +  D_{4y}) \times (D_{5x}  +  D_{5y}) \,\, ,
\end{eqnarray*}
which is used again in its basic operating mode with the Lateral
and Bottom AC ON. This trigger, used for particular data taking
demands or in case of failure of the first plane,   increases the
acceptance angle, at the expenses of a slight  worsening of the
angular resolution. The combination of TRG M2 and High Threshold
again excludes  most of the  protons from the trigger.

The acceptance window  of particles with TRG M1 in
  full containment regime is shown in Table 1, for  Low
Threshold. The spectrum of nuclei extends from hydrogen to iron
in the  energy  interval 10--200 MeV n$^{-1}$.

\vspace{1cm}

\noindent   The flux of particles changes notably along the
orbit. A limit to the acquisition capability of the instrument
has been provided by organizing a system which, in high rate
conditions, enables the detector
 to register events in less
detail. Every 60 seconds the processor in box $D2$  calculates
the rate of particles reaching the detector, and selects one of
the following  acquisition modes:
\begin{enumerate}
\item {\bf Full-Format mode} (counting rate up to 10 Hz).
This mode, in which the whole event topology is recorded,
  is the normal working configuration
outside the Earth's Radiation Belts and in particular out of the
South Atlantic Anomaly (SAA).
 It allows   the measurement of the
 energy released by the
particle in each silicon detector and the storage of this
information.

\item {\bf E$_1$-E$_{tot}$ mode} (counting rate $>$ 10 Hz).  At
high  fluxes it is necessary  to make an optimal use of the mass
memory.  In this acquisition mode, a Second Level trigger, driven
by the processor in $D2$,
 restricts the event acceptance and calculates
  the  energy $E_1$ released in  the  first plane  and the energy $E_{tot}$
deposited in the whole detector by the crossing particle. Instead
of the whole event topology, only $E_1$ and $E_{tot}$ are stored.

\end{enumerate}

\section{Orbit operations}

\subsection{The satellite Resurs-01}

\noindent The class of spacecraft Resurs-01 is designed for
meteorological observations, investigations of Earth natural
resources, and large and small scale ecological monitoring of the
terrestrial environment. Besides these tasks, the spacecrafts are
also utilized for the separation and insertion into orbit
 of small piggy-back space vehicles.

The Resurs-01  is inserted into an almost circular
sun-synchronous orbit, having the parameters shown in Table 2. The
launch of this class of vehicles  takes place from the Baikonur
launch facility in Kazakhstan, by means of a Zenit launcher. The
total mass of the spacecraft at launch, including the separable
micro-satellites,
 is 3200 kg. The mass of the payload alone is 1000 kg.

The service system incorporates a power supply  (providing a
voltage working range from 24 to 34 V), an attitude control and
stabilization system (which maintain the attitude of the
spacecraft
 in three axis), a command
radiolink (designed to control the onboard instrumentation by
radiocommands), a program-time  device (which allows the
spacecraft to be in autonomous operation for 3 days), a
radiotelemetry system, and an onboard time and frequency standard
(which provides stable high frequency and synchronous signals, and
a time mark).

\vspace{0.5cm}

 The instrument NINA is housed into Resurs-01 version n. 4 as shown in
Figure 3. The box $D1$  is mounted on the top side external to
the satellite, in such a way to point always to the zenith during
the flight. The other boxes are located inside the body of the
satellite.

The box $D1$ has two external sensors to measure its temperature.
A  heater keeps it thermal stable when  switched off. The
temperatures of $E$ and $P$, and the stability of the satellite
power supply, are monitored on board Resurs. All these  data are
merged with the satellite telemetry and  sent to the ground.

\subsection{Control in orbit}

\noindent The interaction between the ground stations and NINA
during operations is driven by telecommands, which give the
possibility to activate a total of 24 commands. Some are dedicated
to operations like
 power switching
(ON/OFF),  data transferring,  memory cleaning,  and selection of
single or dual microprocessor model. The others act on the trigger
logic or on the acquisition model, as illustrated previously.
  The
transmission of  specific  telecommands
 can be performed  when the satellite passes
over the ground stations.  A response packet with the status of
the telecommand settings is sent to  Earth  each time a change in
the command buffer  has occurred.

The default telecommand set initializes the acquisition with  TRG
M1, automatic switching between Low and High Threshold enabled,
lateral strips and bottom plane in anticoincidence, and
automatically switching  of the acquisition mode from Full-Format
to ${E_1}$-${E_{tot}}$, according to  the counting rate. However,
it is possible to set any combination of trigger logic, threshold
level, anticoincidence and acquisition mode by telecommand. After
two months of operation, we switched the acquisition to High
Threshold mode, in order to focus our analysis on high Z
particles.

A special onboard device allows pre-programmed combinations of
telecommands, acting automatically in specific points of the
orbit. These combinations permit, for instance, electronic
calibration procedures to be performed over the equatorial
regions (where the counting  rate is low), or to stop the data
acquisition in sectors of the orbit with very high counting rates.

The average volume of data that  NINA transfers from the
satellite to ground is 2 MB day$^{-1}$, corresponding to more
than  20000 events. NINA has a 16 MB mass memory available in the
onboard memory storage. Since the average mass memory occupation
in solar quiet periods and outside the South Atlantic Anomaly  is
around 1.5 MB day$^{-1}$, there is the possibility to accumulate
data for a few days or during solar events, for a subsequent
transmission.

\section{Detector performance in orbit}

The launch of NINA took place on    1998 July 10. The
transmission of its scientific data started on  August 31, after
an initial period needed   for stabilization of the orbit and
overall checks of the satellite functionality. The analysis of
the first  sample of data  received showed that the instrument
 performanced well   and confirmed the
functionality  of the whole system.

During one orbit the satellite has a  day-night cycle according
to its position with respect to the Earth and the Sun. Two of the
34 housekeeping data available on NINA
 give internal temperatures sampled at two different heights inside the detector.
We measured the behavior of the two temperatures in orbit.  The
excursion of their values between light and shadow was less than 1
degree, as required in the construction phase.

Important information about the status of the detector are
provided  by
 the ratemeters, which are also  part of the housekeeping sector. These are
indicators
 of the particle flux impinging on different planes of box $D1$, which is a
function of  the orbit of the satellite. Low  and high flux
ratemeters are implemented at different heights inside the
telescope. In case of intense flux, the high ratemeters section
provides information while the low ratemeters may saturate.

Figure 4 shows the behavior of the low  ratemeter implemented on
plane n.6 during one typical orbit. The counting rate  is given in
hertz, with the saturation value at about 420 Hz.
 From the picture one can easily
follow
the path of the satellite through the different regions of the Earth's
magnetosphere. In particular, it can be seen how the flux increases at the Poles with respect
 to the Equator,
because at high latitudes the terrestrial magnetic field does not
effectively prevent  low energy particles from approaching the
Earth. The spikes visible near the Poles are due to low energy
electrons which fill the Outer Radiation Belt. In  the South
Atlantic Anomaly the magnetic field has a local minimum and thus
the low energy proton  flux reaches very high levels. This is
clearly evident by the saturation of the ratemeter counter.

We have examined the stability of some important parameters of
the detector with time, during the first 6 months of life of the
detector. The pedestal values remained stable within 1 ADC
channel from October 1998 until March 1999. The same holds for
the voltages, the threshold values, the temperatures.

\section{Galactic Cosmic Ray measurements}

As mentioned before, after the first two months of operation
NINA's activity was focused on the detection of particles heavier
than hydrogen. In section 6.1 we discuss the track selection
algorithm for Z$>$1 particles, that we utilized to calculate the
GCR fluxes of  $^4$He, $^{12}$C, and  $^{16}$O, which are  shown
in section 6.3. The algorithm of isotope identification  and the
performance of mass discrimination of NINA in orbit are
presented  in section 6.2.

\subsection{The track selection algorithm}

The optimal performance of NINA
 in terms of charge, mass and energy
determination is achieved by requesting the full containment of a
particle inside the detector, using the Lateral and Bottom
Anticoincidence System as a veto. In order to reject upward
moving particles, tracks accompanied by nuclear interactions, and
 events consisting of two and more tracks, an off-line track
selection algorithm  for the data analysis is needed.

The selection algorithm for nuclei with Z$>$1, implemented for the analysis of NINA flight data,  applies  six rejection criteria:

\begin{enumerate}

\item Real particles moving downwards and stopping  inside the detector have an energy deposition that increases
along the track.
It is natural, therefore,  to  request tracks to deposit in each view a quantity of energy greater than
in the previous one multiplied by a constant $K_1$, which takes into account the energy fluctuations.  If

\[ E(i) < K_1 \times E(i-1), \]

for any  $i$ in the range from the second crossed view to the one with the maximum deposit of energy,
the event is rejected.

\item In order to clean the data sample from particles with nuclear interactions,  two energies for each track are calculated:

- E$_{track}$ = sum of the energies released  by the particle from the first hit view to the view following the one with the maximum deposit of energy;

 - E$_{residual}$ = total amount of energy left in all the remaining layers.

The two energies are compared and events with \[ E_{residual} > K_2 \times E_{track}, \] where $K_2$ is a parameter to be optimized, are rejected.

\item Double tracks are eliminated estimating two energies for each crossed view $i$ along the particle path:

- E$_{cluster}(i)$ = sum of energy released in the strip with the maximum deposit of energy and in the two nearest strips;

- E$_{noise}(i)$ =  sum of the energy released in the other strips of the silicon layer.

If \[E_{noise}(i) > K_3 \times E_{cluster}(i), \]

for any of the $i$ crossed views and where $K_3$ has to be fixed,  the event is rejected.

\item Events with the maximum deposit of energy in the first view are rejected. This criterion, together with the
condition n. 1, selects downwards moving particles.

\item Events where the maximum energy release in the X view and in the Y view
are not in  the same or between consecutive detector planes are
rejected. This request helps filtering double tracks.

\item In order to reduce the number of particles which leave the detector through the space between planes,
 events which release the {\it maximum} of the energy deposit per silicon layer in the strips 2 or 15, for any of the crossed layers,   are  rejected.

 \end{enumerate}

To apply this algorithm  to the data collected in orbit it was
necessary to choose the $K_1$, $K_2$, and $K_3$ coefficients in
such a way to efficiently clean the data sample from the
background, minimizing at the same time the number of good events
rejected.

In order to optimize these values we utilized  samples of
different types of particles,  obtained from a  beam test session
at GSI in 1997 \cite{nina2}, as well as flight data.

If we define as efficiency $\epsilon$ the value $\epsilon = (1 -
\frac{N.\: rejected \:good \:events}{N. \:good \:events})$, the
best  optimization of the $K_1$, $K_2$, and $K_3$ values that we
achieved ($K_1=0.7$, $K_2=0.01$ and $K_3=0.01$) determined an
efficiency equal to $\epsilon =0.975\pm0.003$ for all particles
with Z$>$1.

\subsection{Isotope identification}

\noindent  Charge and mass identification procedures may be applied to the events which
survive the track selection algorithm.

The mass M and the charge Z of the particles
are calculated in parallel by two methods, in
order to have a more precise particle recognition:
\begin{enumerate}

\item[a)] the method of the residual range
\cite{sampex,Hasebe,massadur,nina2}.

In this method, the charge Z is estimated by means of the product
${E_1 \times E_{tot}}$. Here $E_1$ is the energy released by the
particles in the first silicon detector (two layers) of the tower
NINA, and  $E_{tot}$ is the total energy released in the whole
instrument. Figure  5 shows  the $E_1 \,\,\, vs \,\,\, E_{tot}$
curves of  particles resulting  from the fragmentation of
$^{12}$C  by means of a polyethylene target, obtained during a
beam test of NINA \cite{nina2}. The nuclear families lie on
different hyperbolas   $E_1 \times E_{tot} = k(Z^2)$.

Once the charge Z is identified by its $E_1 - E_{tot}$ hyperbole,
the mass of the particle is  evaluated by applying the following
formula:
\begin{equation}
M = \left( {{a{( E_{tot} ^b - (E_{tot}  - {\Delta}E )^b)} }\over{Z^2 {\Delta}x}}\right)^
{{1}\over{b - 1}} \; \;   ,
\label{eq4-3}
\end{equation}

\noindent where $\Delta E$ is the energy lost by the particle in a
thickness $\Delta x$ measured starting from the first plane, the
parameter $a$ is a constant depending on the medium and $b$ has a
value between 1.5 and 1.8 in NINA's energy range. A precise
evaluation of such parameters for each atomic species   has been
obtained both from real and  simulated data.

\item[b)] the method of the approximation to the Bethe-Bloch theoretical curve.

With this second method we estimate the mass M and charge Z of the particle by minimizing the following
$ \chi ^2$ quantity :
\begin{equation}
\chi  ^2 = \sum ^{N} _{i=1} \left ( W _i  \left( \Delta E ^{real} _ {i} - \Delta E ^{theor}
_{i} \right ) \right ) ^2 \,\, ,
\end{equation}

where ${ \Delta E ^{real}_ {i}}$ is the energy released by the
particle in the i-th view,  ${ \Delta E ^{theor}_ {i}}$ is the corresponding
expected value, $W _i$  is the weight for every difference
${W _i = {1 \over {\Delta E ^{real}_ {i}}}}$, and the
 sum is extended to the $N$ silicon layers
activated by the particle, excluding the last one where the particle stops and the fluctuations of the energy deposits are generally very big.

In order to build such a function, it is necessary to follow step
by step the particle's path, calculating the scattering angles at
every  layer. This method takes into account also the energy
losses in  dead layers, thus preventing   systematic shifts on
the reconstructed masses.
\end{enumerate}

For a complete  rejection of the background, only particles with
the same final identification given by the two methods are
selected. Finally, a cross-check between the real range of the
particle in the detector and the expected value according to
simulation is a  consistency test for the event.

\vspace{0.5cm}

In Figure 6 the reconstructed   masses   using eq. 1 for   helium
isotopes  detected in orbit are compared to the ones obtained
 from the  test-beam data;
the  sample of particles in orbit has been selected during
passages over the polar caps, and in period of  quiet solar
activity. The picture shows that the  mass resolution of NINA in
flight  is   in  good agreement with the measurements performed
at GSI.

\subsection{Determination of Fluxes}

The analysis presented in this section refers to particles
registered by NINA in the solar quiet period December 1998-March
1999, detected in High Threshold mode and TRG M1 acquisition. In
order to select a sample of pure low energy (E$ > $10 MeV/n)
primary cosmic rays, and avoid the distortions induced by the
Earth magnetic field, only particles registered at a value of
L-shell$>$6 (L geomagnetic shell) were chosen.

In order to estimate the cosmic ray  fluxes it is also necessary
to know the geometric factor of the instrument as a function of
energy, and the exposure time in orbit. A correction factor can
then be applied to account for energy-loss  in the aluminum
window.

The geometric factor of the instrument
was calculated by means of Monte Carlo simulations based on
the CERN-GEANT code \cite{geant}.
Each simulated track underwent the trigger conditions (TRG M1 or TRG M2),
 with the Lateral and Bottom Anticoincidences ON, both for Low and High Threshold mode.
Figure 7  presents the geometric factor G of NINA for $^4$He,
$^{12}$C, and $^{16}$O  in  High Threshold mode over   the energy
intervals defined by the trigger and the selection conditions
explained before.

\vspace{0.5cm}

The incoming  energy of the particles was reconstructed  by an
iterative algorithm which is based on the Bethe-Bloch formula.
The algorithm works this way: as a first step, the total energy
E$_{tot}$ is taken as initial energy  E$_{in}$ of the particle;
with this value of initial energy, all energies deposited in
every layer along the particle track (namely the aluminum window,
the silicon detectors and the inter-gap volumes of nitrogen) are
calculated, and the expected  value of the total energy deposited
in the silicon E$_{tot}^{exp}$  estimated.

The value of E$_{tot}^{exp}$  is then compared with E$_{tot}$. If
their difference is greater than    0.1 MeV we  define a new
initial energy as E$_{in}$ = E$_{tot}^{exp}$ + E$_{step}$, where
E$_{step}$ is an incremental step energy  fixed according to the
precision that we want to reach,
 and perform a new iteration.
When finally the condition  \[ E_{tot} - E_{tot}^{exp} \leq    0.1 \, MeV \]
 is fulfilled, the algorithm stops and
  the initial energy of the particle is identified.

\vspace{0.5cm}

The differential energy spectra  were then determined by the
following formula:

\[ Flux(E)= \frac{\Delta N(E)}{ T \: \epsilon \: G(E) \: \Delta E }, \]

where $\Delta N(E)$ is the number of detected particles with energy between $E$ and $E+\Delta E$,
$T$ is the  exposure time  in orbit for the period under consideration, $\epsilon$
is efficiency of track selection discussed above (0.975 for nuclei with Z$>$1), $G(E)$ is the average value of the geometrical factor between $E$ and $E+\Delta E$, and $\Delta E$ is the energy bin chosen to plot the flux.

Figures 8, 9, and 10 present respectively the differential energy
spectra for $^4$He,  $^{12}$C, and  $^{16}$O, measured by NINA in
the solar quiet period December 1998-March 1999. Errors due to
energy resolution are negligible since the energy resolution for
NINA is better than 1 MeV n$^{-1}$, which is much less than the
width of the energy bins in the flux plots.

In Figure 8 NINA flux of $^4$He is plotted together with the
results from  the mission SIS on ACE, about in the same period of
observation. Data from SIS on ACE \cite{webace}  belong to a
cycle of 27 days
 from 1999  February 6 to 1999  March 4, and are the sum of $^3$He and $^4$He; errors are statistic plus systematic.

There is a general agreement among  the two sets of results. However there are differences between
the results of NINA and SIS, which can be attributed to the different  time period (since the flux of helium
is known to change significantly  over the months) and to the fact that the SIS flux include also $^3$He.

In Figures 9 and 10  the NINA  differential energy spectra of
respectively $^{12}$C and $^{16}$O
 are plotted together with results from  the missions SIS and CRIS on ACE, both
referred to the period  1999  February 6 to 1999  March 4 \cite{webace}.
The  fluxes measured on board NINA are in very good agreement with the ones on ACE.

\section{Conclusions}

The space telescope NINA, launched in 1998,
is a silicon detector devoted to the study of cosmic rays of galactic, solar
and anomalous origin in the energy range 10--200 MeV n$^{-1}$ at 1 AU.
It is capable of nuclear identification up to iron and isotopic discrimination up
to nitrogen, allowing  important space physics issues, such as the
composition and energy spectra of cosmic ray particles, to be addressed.

The first months of  data analysis confirmed that the instrument is working properly in
space; the overall performances of the detector
are in good agreement with expectations and,
in particular,  the mass resolution capability reached by NINA in space
reproduces the one obtained in a beam test session.

The energy spectra of galactic $^4$He, $^{12}$C, and $^{16}$O
measured by NINA at 1 AU have been determined. The analysis of the
galactic ratio $^3$He/$^4$He together with the abundance ratio of
the isotopes of hydrogen is in progress, as well as the study of
the composition and energy spectra of SEP and ACR particles.

The measurements performed by   NINA  are  important  in view of
the second  mission NINA-2   which will complement the
observations of NINA extending its lifetime  to cover a complete
solar cycle. The addition of  PAMELA \cite{pamela1,pamela}  will
allow the extension of cosmic ray observations  to energies
greater  than 200 GeV.

\acknowledgments{
We acknowledge  the Laben  company  (Italy)  for the realization of $D1$ and $D2$ parts of
the detector as well as VNIIEM  (Russia) for parts $E$, $P$ and especially for  the
assembly and space operations on the Resurs satellite.

Finally we acknowledge  the Russian Foundation of Base Research, grant 99-02-16274,
who partially supported the Russian Institutions for this work.}

\begin{table}
\caption{Energy windows  for the most abundant particles fully
contained in the  detector NINA, with TRG M1 and
 acquisition with Low   Threshold.\label{thre}}
\vspace{0.5cm}
\begin{tabular}{ccc}
\tableline
 Particle & Z &  Energy window (MeV n$^{-1}$)\\ \tableline
\tableline
 $^1$H  &  1&   10 - 48 \\
 $^4$He& 2   &9 -   50   \\
 $^7$Li &3  &11 - 54  \\
 $^9$Be &4&13 - 65   \\
 $^{11}$B &5 &15 - 75  \\
 $^{12}$C&6 &17 - 90  \\
 $^{14}$N   &7 &18 - 95  \\
 $^{16}$O &8&   20 -103  \\
 $^{19}$F &9 &21 - 107 \\
 $^{20}$Ne& 10 & 23 -117 \\
 $^{28}$Si& 14 & 28 - 142 \\
 $^{40}$Ca& 20 & 39 - 175 \\
 $^{56}$Fe& 26 & 58 - 195 \\
\tableline
\end{tabular}
\end{table}

\clearpage
\begin{table}
\caption{Parameters of the Resurs-01 n.4 satellite.}
\vspace{0.5cm}
\begin{tabular}{cc}
\tableline
 Characteristic & Value \\ \tableline
\tableline
Orbit inclination (deg) & 98.75 \\
 Orbit period (min) & 101.31 \\
 Eccentricity & 1.12$\times$10$^{-3}$ \\
3 axis stabilization accuracy  (deg) & 1.0  \\
 Average orbit altitude   (km)  & 840 \\
\tableline
\end{tabular}
\end{table}

\clearpage

\begin{figure}
 \figurenum{1}
\epsscale{0.8}
\plotone{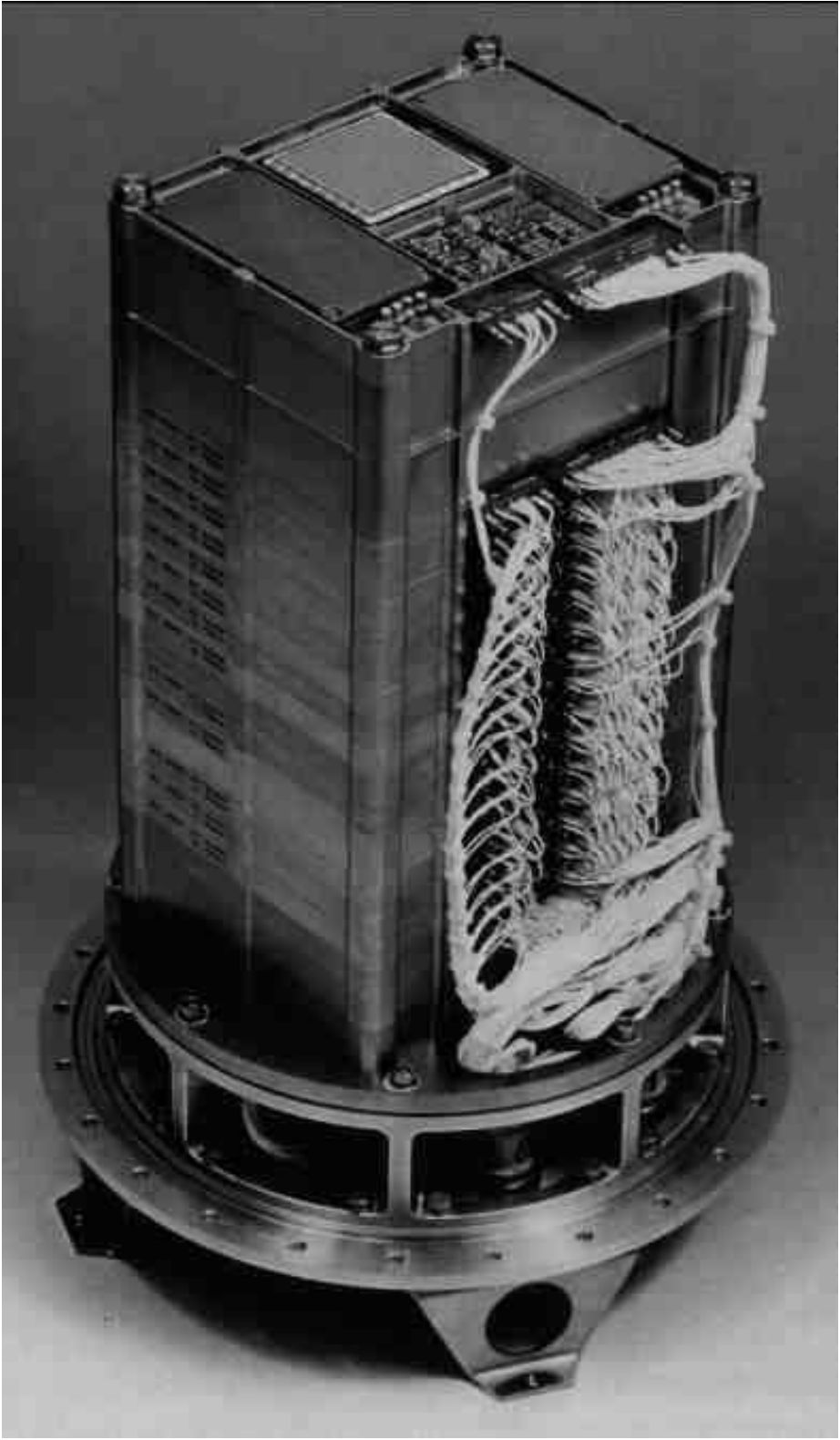}
   \caption{Photograph of the internal structure of  box $D1$.}
\end{figure}

\begin{figure}
 \figurenum{2}
\epsscale{1} \plotone{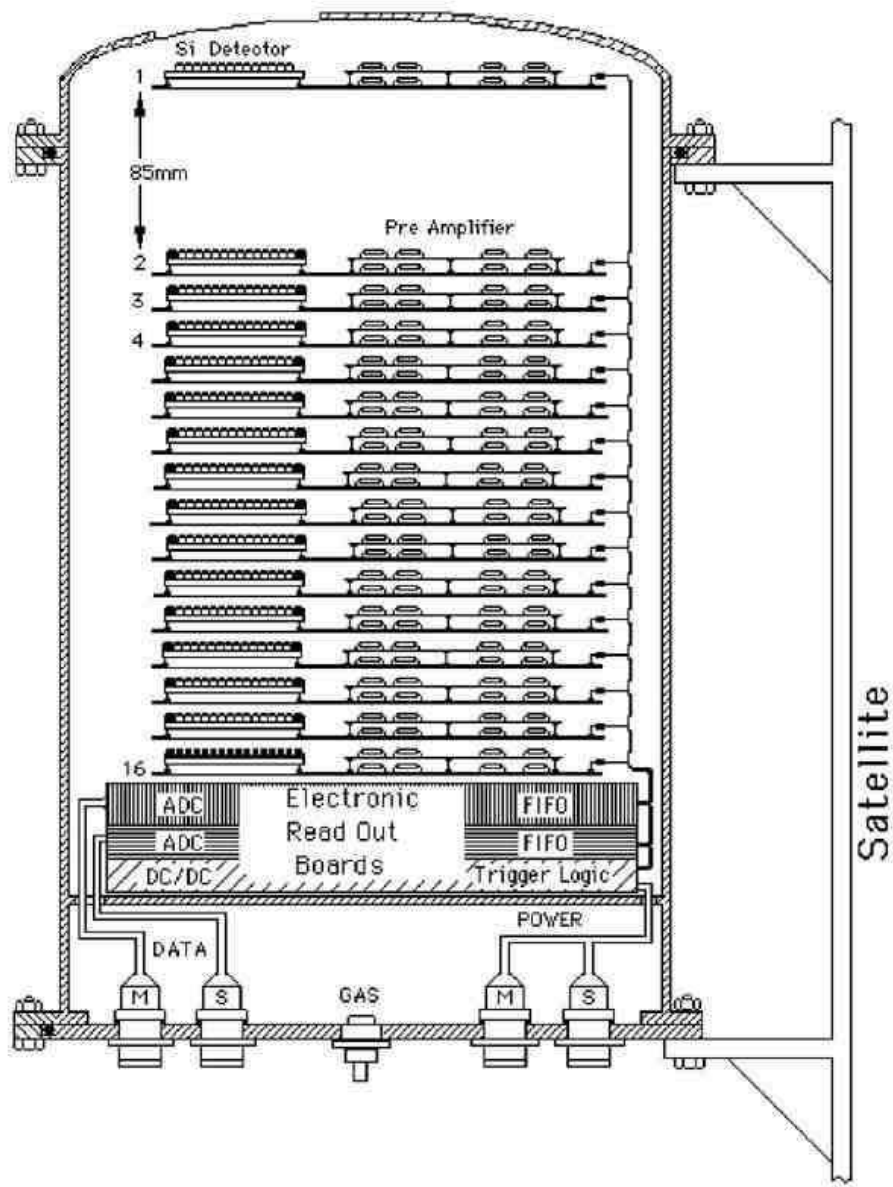}
   \caption{Sketch of the internal structure of  box $D1$.}
\end{figure}

\begin{figure}
   \figurenum{3}
\epsscale{1} \plotone{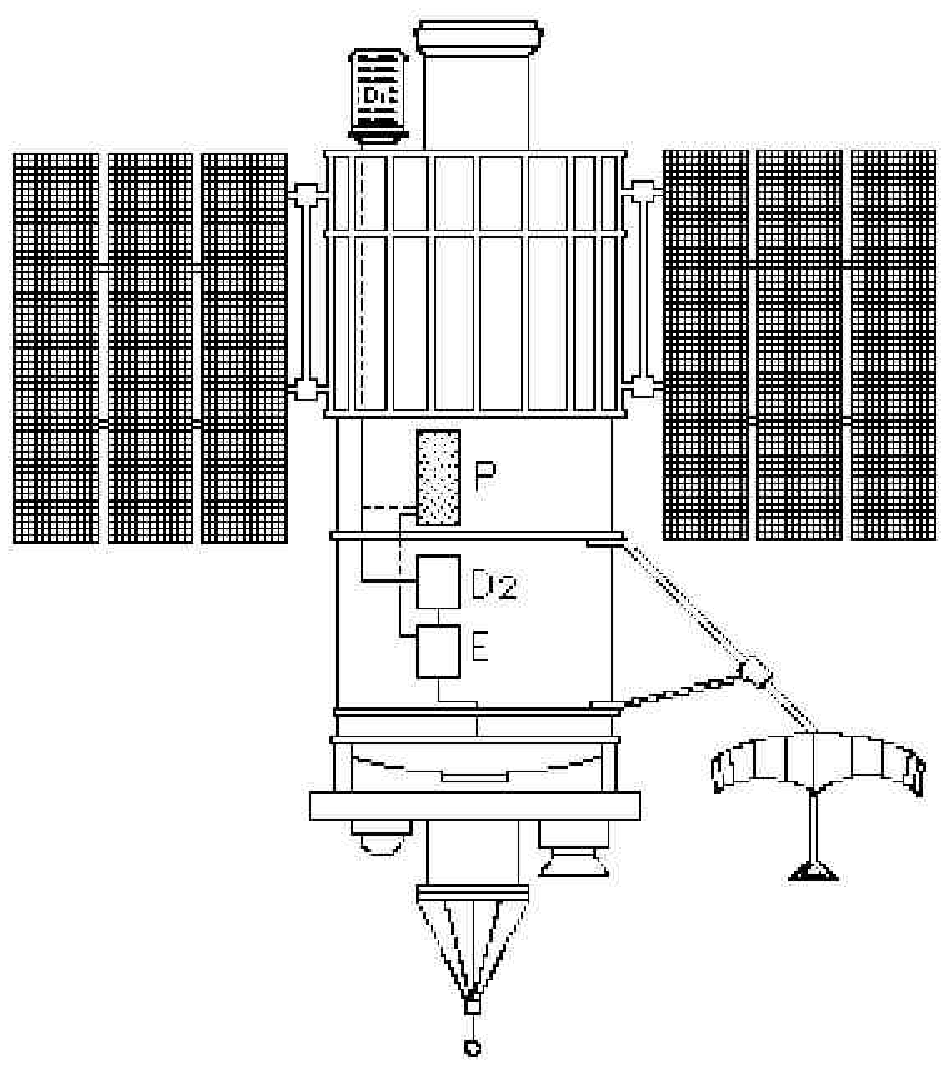}
 \caption{Location of the various subsystems of NINA -
$D1$, $D2$, $E$ and $P$ -
 on the satellite Resurs. }
\end{figure}

\begin{figure}
  \figurenum{4}
\epsscale{1} \plotone{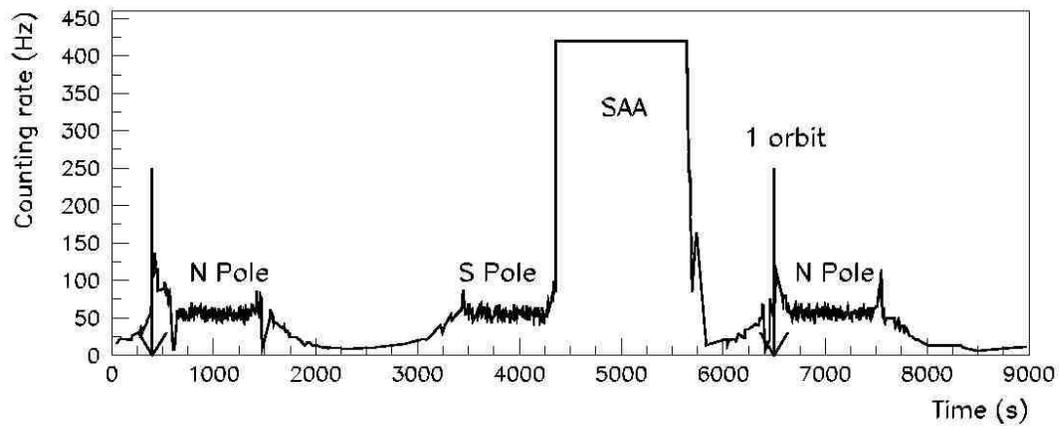}
 \caption{Low ratemeter counting rate (at plane 6)
 as a function of time.
 The two black vertical arrows define 1 satellite orbit.
 }

\end{figure}

\begin{figure}
   \figurenum{5}
\epsscale{1} \plotone{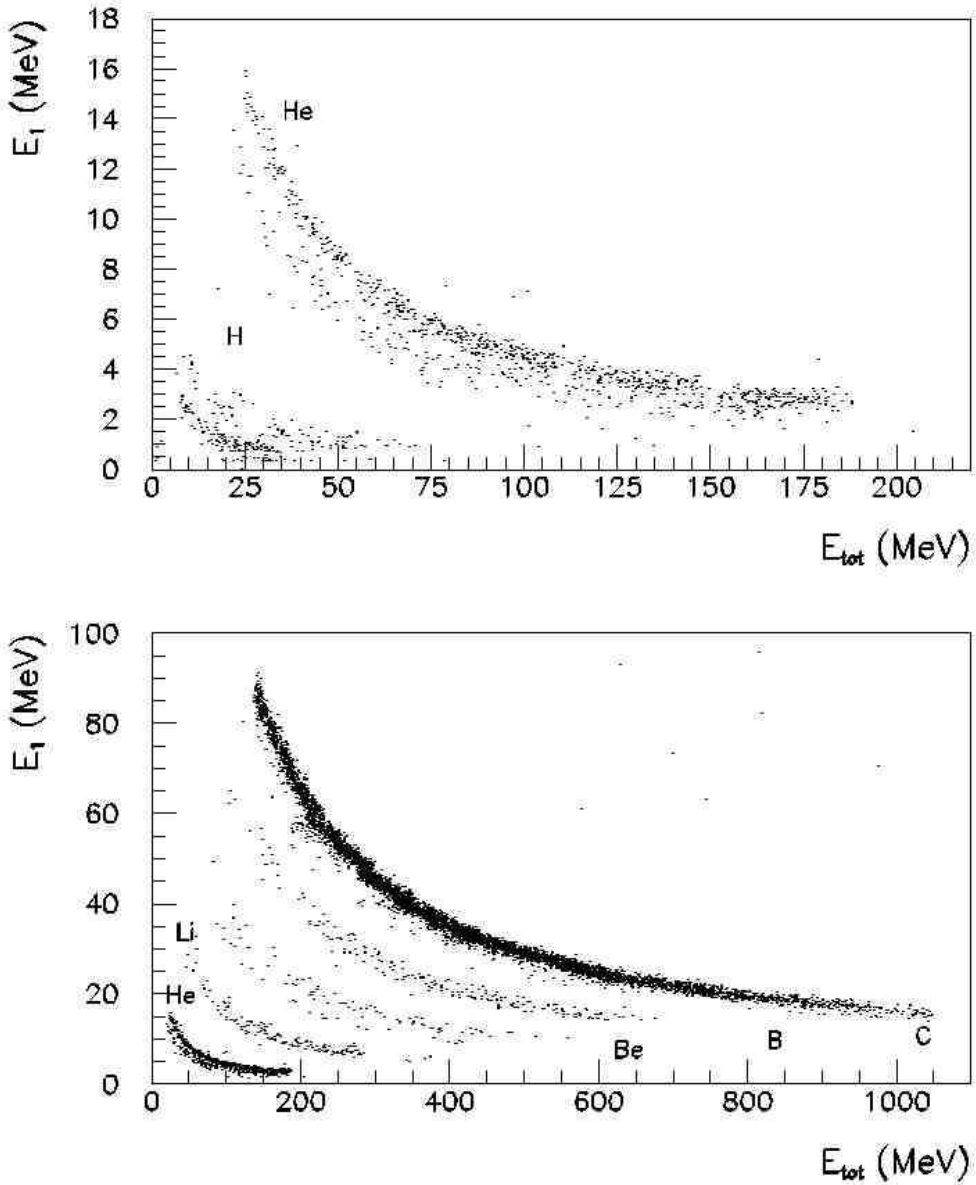}
 \caption{Distribution of the energy released in the
first plane ($E_1$)  and the total energy
($E_{tot}$) detected for  particles
produced in the fragmentation of
$^{12}$C at GSI (1997). }

\end{figure}

\begin{figure}

\figurenum{6} \epsscale{1} \plotone{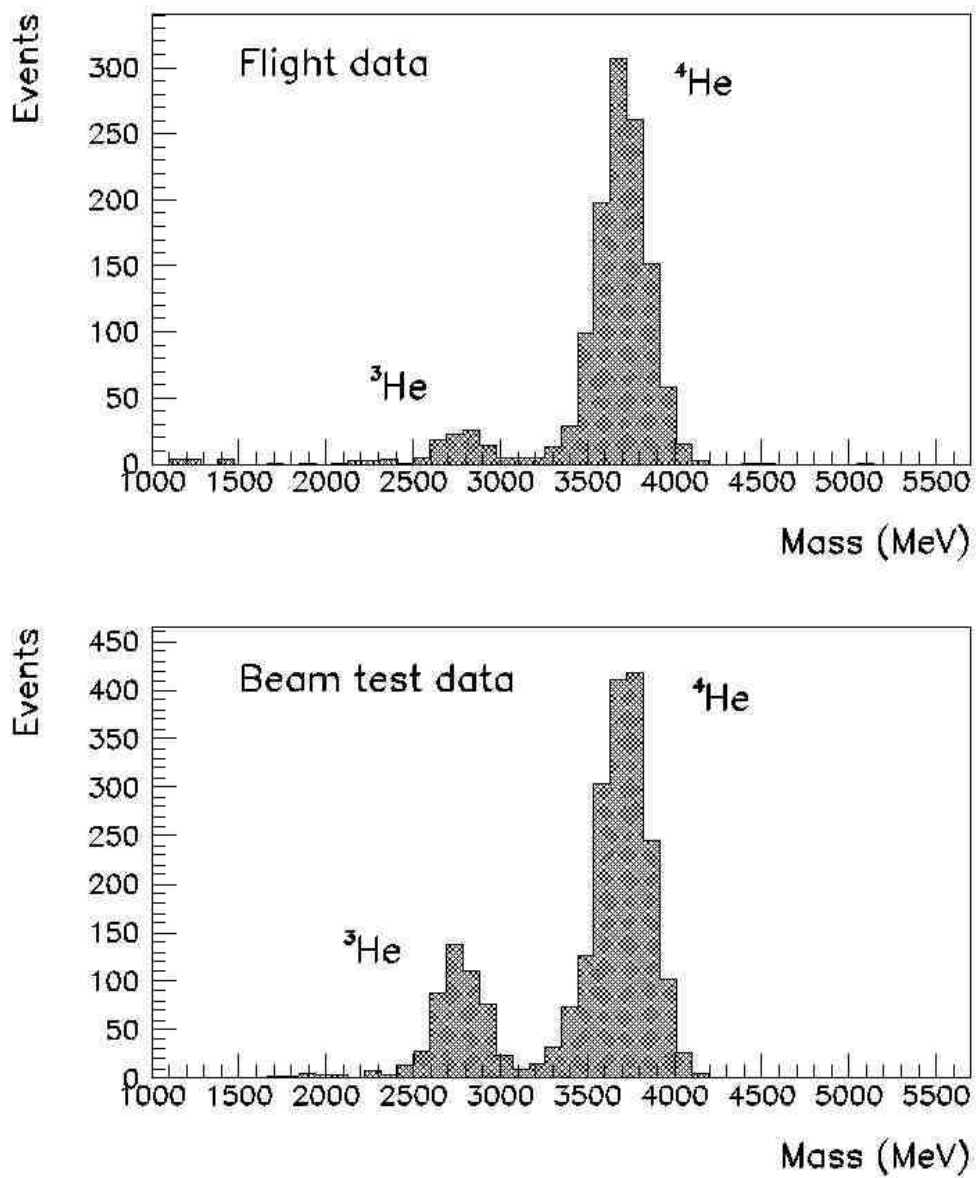}
 \caption{Mass distribution, as given by eq. 1,   of a sample of helium  isotopes
collected by NINA in orbit (top) and during the beam test session at GSI (bottom).}

 \end{figure}

\begin{figure}
  \figurenum{7}
\epsscale{1} \plotone{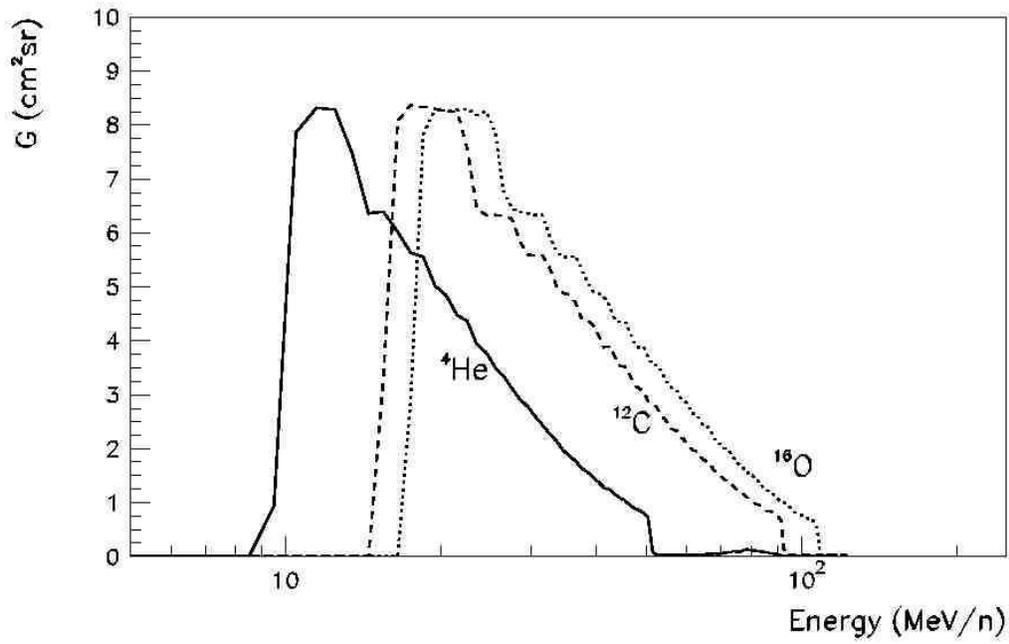}
    \caption{Geometric factor G of NINA  for $^4$He, $^{12}$C, and $^{16}$O  in High Threshold mode.}
\end{figure}

\begin{figure}
  \figurenum{8}
\epsscale{0.7} \plotone{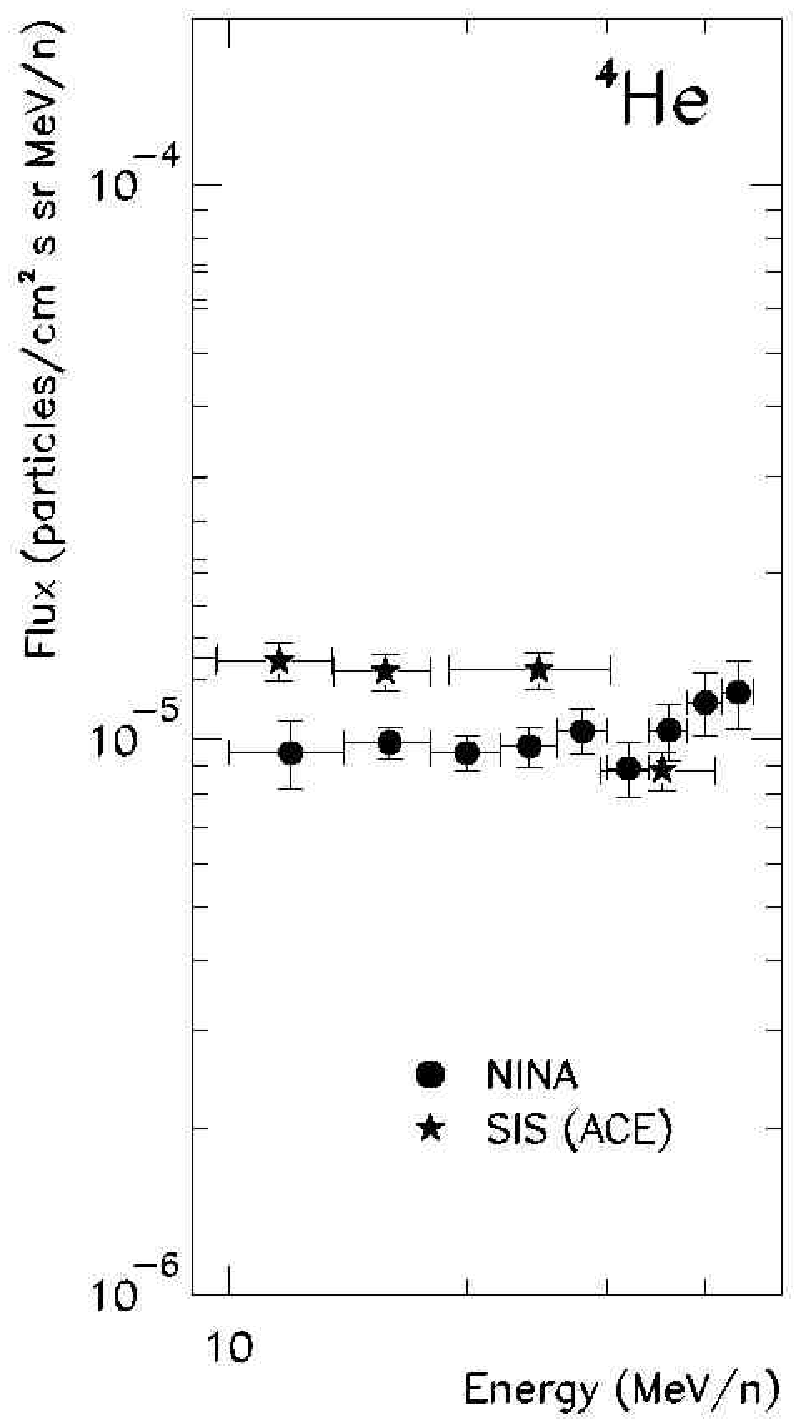}
    \caption{Differential energy spectrum for $^4$He in the solar quiet period
December 1998-March 1999 measured by NINA, together with data of SIS on board ACE.}
\end{figure}

\begin{figure}
  \figurenum{9}
\epsscale{0.7} \plotone{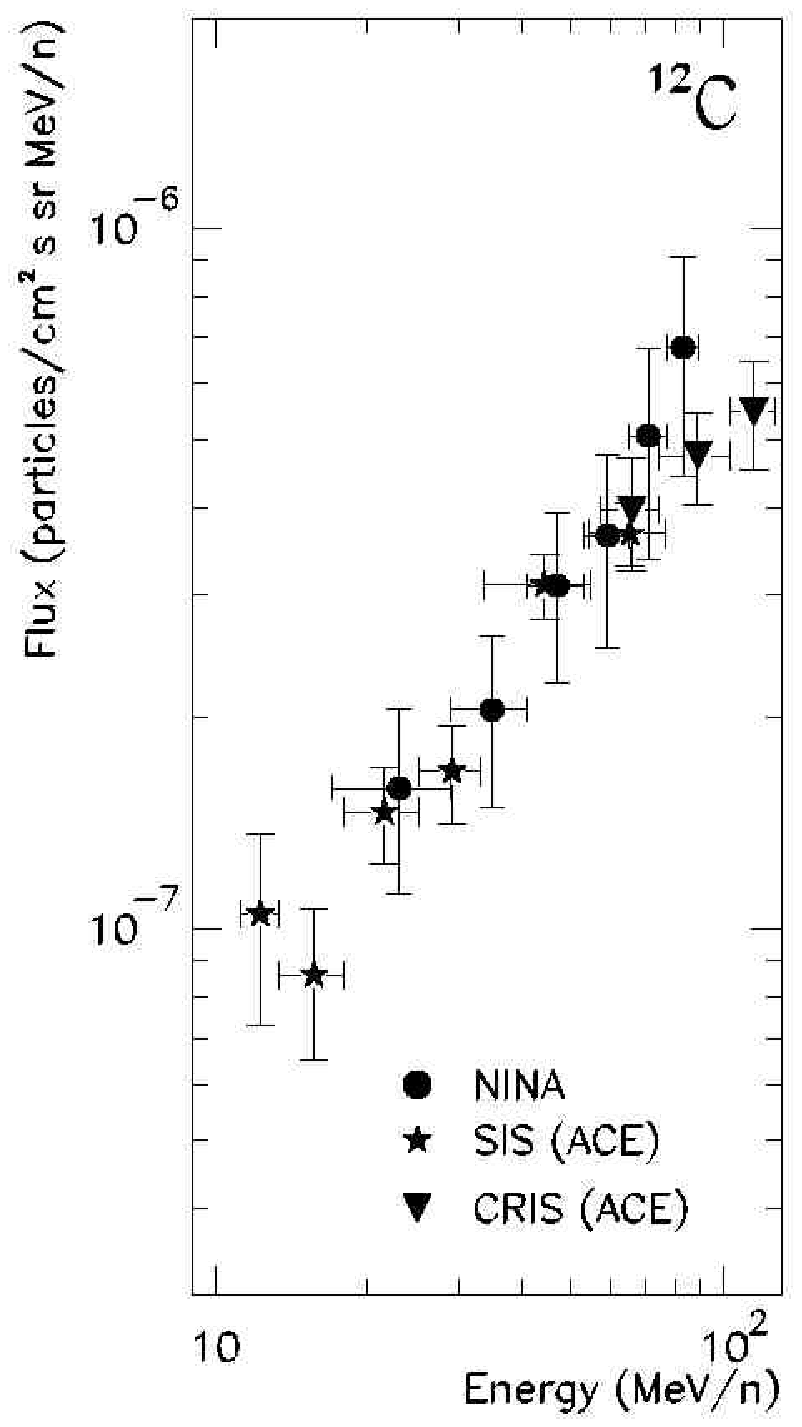}
    \caption{Differential energy spectrum for $^{12}$C in the solar quiet period
December 1998-March 1999 measured by NINA, together with data of SIS and CRIS on board ACE.}
\end{figure}

\begin{figure}
  \figurenum{10}
\epsscale{0.7} \plotone{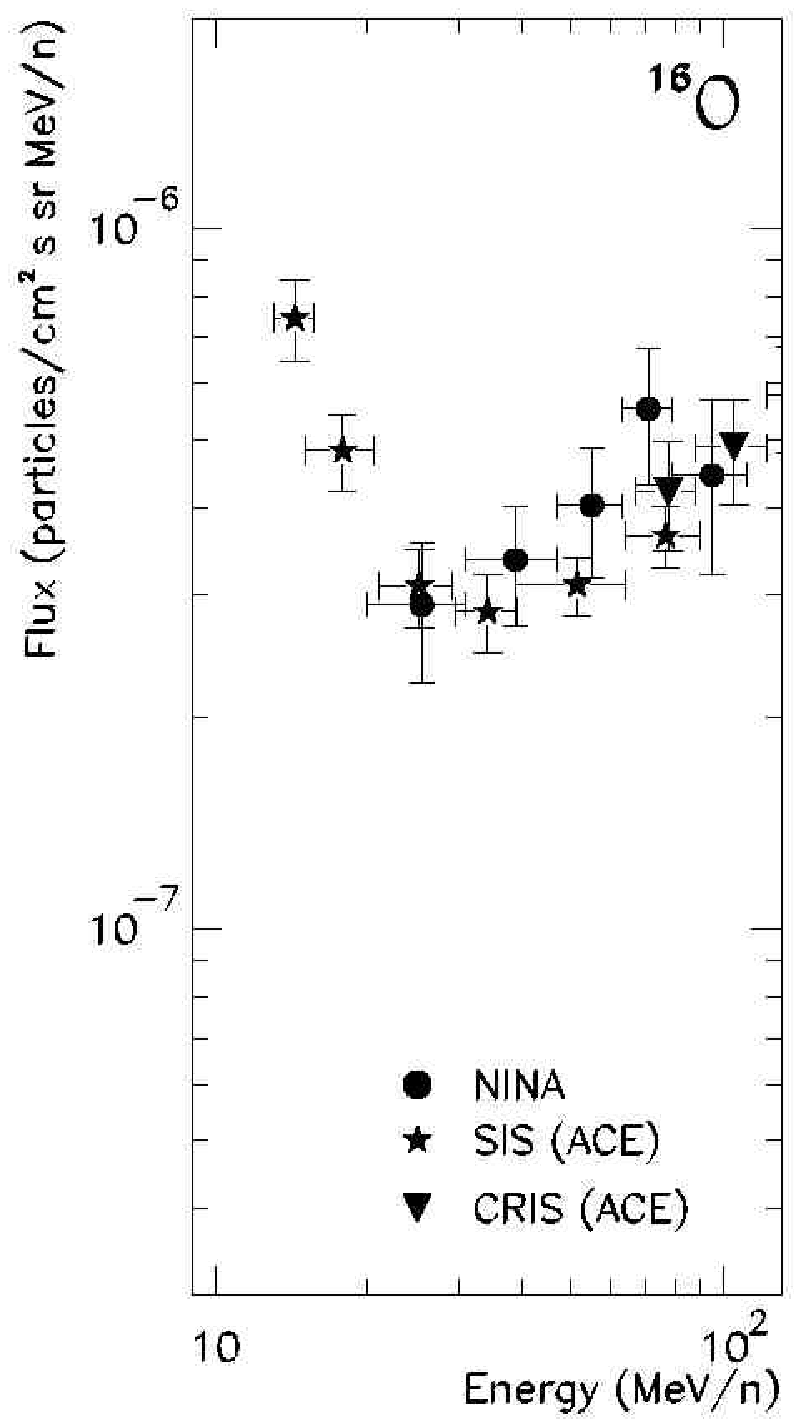}
    \caption{Differential energy spectrum for $^{16}$O in the solar quiet period
December 1998-March 1999 measured by NINA, together with data of SIS and CRIS on board ACE.}
\end{figure}


\newpage

\baselineskip=24pt
\begin{thebibliography}{99}

\bibitem[Adriani et al. 1997]
{pamela} Adriani, O., et al. 1997,  Proc. 25 Int. Cosmic-Ray Conf. (Durban),  5, 49.
\bibitem[ACE Home Page]{webace} ACE Home Page, Online Data, Level 2, http://www.srl.caltech.edu/ACE/ASC/level2.
\bibitem[Bakaldin et al. 1997]{nina1}  Bakaldin, A., et al. 1997,  Astrop. Phys., 8, 109.
\bibitem[Baker et al. 1993]{sampex}  Baker, D. N., Mason, G. M., Figueroa, O., Colon, G., Watzin, J. G. \&
Aleman, R. M.   1993,   IEEE Trans. on Geoscience and Remote
Sensing, 31-3, 531.
\bibitem[Barbiellini et al. 1996]{pal5} Barbiellini, G., et al.  1996,  \aap, 309, L15.
\bibitem[Basini et al. 1999]{pal2} Basini, G.,  et al. 1999,  Proc. 26 Int. Cosmic-Ray Conf. (Salt Lake City), 3, 85.
\bibitem[Bidoli et al. 1999]{nina2}  Bidoli, V., et al.  1999, Nucl. Instr. Methods Phys. Res., A 424, 414.
\bibitem[Boezio et al. 1997] {pal4}Boezio, M.,  et al.  1997,  \apj,
487, 415.
\bibitem[Boezio et al. 1999]{pal3}  Boezio, M., et al. 1999,  Proc. 26 Int. Cosmic-Ray Conf. (Salt Lake City), 3, 57.
\bibitem[Boezio et al. 2000]{pal1} Boezio, M., et al.  2000,  to appear on \apj,  531.
\bibitem[Brun et al. 1994]{geant} Brun, R., et al. 1994, ''Detector Description and Simulation Tool'', CERN program library.
\bibitem[Casolino et al. 1999a]{slc1}Casolino,   M., et al. 1999a,  Proc. 26 Int. Cosmic-Ray Conf. (Salt Lake City), 5, 108.
\bibitem[Casolino et al. 1999b]{marcoicrc} Casolino,   M., et al. 1999b,  Proc. 26 Int. Cosmic-Ray Conf. (Salt Lake City),  5, 136.
\bibitem[Fisk$,$ Kozlovsky \& Ramaty 1974]{fisk}  Fisk, L. A., Kozlovsky, B. \& Ramaty, R.  1974, \apj,   190, L35.
\bibitem[Golden  et al. 1994]{pal8}Golden, R. L., et al.  1994, \apj ,
436, 769.
\bibitem[Golden et al. 1996]{pal7} Golden, R. L.,  et al.  1996,  \apj,
457, L103.
\bibitem[Gosling 1993]{cliver} Gosling, J. T.  1993, \jgr,  98, 18949.
\bibitem[Hasebe et al. 1993]{Hasebe} Hasebe,  N., et al. 1993,  Nucl. Instr. Methods Phys. Res., A 325, 335.
\bibitem[Hof et al. 1996]{pal6}  Hof, M., et al. 1996, \apj,
467, L33.
\bibitem[Klecker 1995]{klecker} Klecker,  B. 1995, Space Science Reviews, 72 , 419.
\bibitem[Klecker et al. 1998]{klecker2} Klecker, B., et al. 1998,  Space Science Reviews, 83, 259.
\bibitem[Pesses$,$  Jokipii \& Eichler 1981]{pesses} Pesses, M. E., Jokipii, J. R. \& Eichler, D.   1981, \apj,  246, L85.
\bibitem[Reames 1990]{ream1}  Reames,  D. V.  1990, \apjs,   73,  235.
\bibitem[Reames 1993]{ream2}  Reames,  D. V.  1993, Adv. Space Res., 13, n.9, 331.
\bibitem[Reames 1995]{ream3} Reames, D. V.  1995 , Rev. Geophys. Suppl., 33, 585.
\bibitem[Reames 1998]{ream4}  Reames,  D. V.  1998, Space Science Reviews, 85, 327.
\bibitem[Simpson 1995]{simp0} Simpson,  J. 1995, Adv. Space Res., 16, n.9, 135.
\bibitem[Sparvoli et al. 1997]{massadur} Sparvoli,  R., et al.  1997, Proc. 25 Int. Cosmic-Ray Conf. (Durban),  2, 181.
\bibitem[Sparvoli et al. 2000]{smi} Sparvoli,  R., et al. 2000,  Nuclear Physics B, 85, 28.
\bibitem[The Pamela Coll. 1999]{pamela1} The Pamela Collaboration  1999, Proc. 26 Int. Cosmic-Ray Conf. (Salt Lake City),  5, 96.
\bibitem[Wiedenbeck \& Greiner 1980]{gal1}  Wiedenbeck, M. E. \& Greiner,  D. E.   1980, \apj,  239, L139.

\end{thebibliography}
\end{document}